# Fabrication and electrical integration of robust carbon nanotube micropillars by self-directed elastocapillary densification

Michaël F.L. De Volder[1,2,3], Sei Jin Park[1], Sameh H. Tawfick[1], Daniel O. Vidaud[1], and A. John Hart[1,*]

[1]Department of Mechanical Engineering
University of Michigan
2350 Hayward Street
Ann Arbor, MI 48109 USA
* ajohnh@umich.edu

[2]IMEC, Kapeldreef 75, 3001 Heverlee, Belgium
[3]Department of Mechanical Engineering, Katholieke Universiteit Leuven, 3001 Leuven, Belgium

**Abstract**

Vertically-aligned carbon nanotube (CNT) "forests" microstructures fabricated by chemical vapor deposition (CVD) using patterned catalyst films typically have a low CNT density per unit area. As a result, CNT forests have poor bulk properties and are too fragile for integration with microfabrication processing. We introduce a new self-directed capillary densification method where a liquid is controllably condensed onto and evaporated from the CNT forests. Compared to prior approaches, where the substrate with CNTs is immersed in a liquid, our condensation approach gives significantly more uniform structures and enables precise control of the CNT packing density. We present a set of design rules and parametric studies of CNT micropillar densification by self-directed capillary action, and show that self-directed capillary densification enhances the Young's modulus and electrical conductivity of CNT micropillars by more than three orders of magnitude. Owing to the outstanding properties of CNTs, this scalable process will be useful for the integration of CNTs as functional material in microfabricated devices for mechanical, electrical, thermal, and biomedical applications.

**1. Introduction**

While the outstanding properties of carbon nanotubes (CNTs) [1, 2] have generally been verified by characterization of individual CNTs [3, 4], the collective electrical [5], mechanical [6], and thermal [7] properties of CNT assemblies have largely not satisfied predictions based on scaling laws. This shortcoming is mainly due to challenges in fabricating CNT assemblies, namely controlling the diameter, packing density, and straightness of CNTs, and establishing robust electrical and mechanical contact to the CNTs. Further, integration of CNT assemblies as functional material for uses such as microelectronic interconnects and thermal interfaces [5, 7] requires parallel placement of large numbers of assemblies under conditions compatible with microfabrication.

The most common approach to fabricating CNT assemblies for use in microsystems is growth of vertically aligned CNT "forests" from a catalyst layer by chemical vapor deposition (CVD) [8]. This approach enables vertical self-organization of CNTs on a silicon substrate, and this mechanically favorable configuration facilitates growth of films comprising parallel CNTs whose length can reach up to several millimeters. Further, patterning of the catalyst layer by photolithography enables fabrication of CNT forest microstructures having virtually any cross-sectional shape. The lateral dimensions of CNT forest microstructures are limited by the lithography resolution, and their aspect ratio (forest height) is limited by the growth process and catalyst lifetime [9]. Nevertheless, the dimensional limits of CNT forests grown by CVD are comparable to vertical microstructures made by top-down methods such as DRIE (Deep Reactive

Ion Etching) [10]. Even at 10% of the ideal stiffness and strength of individual CNTs, the mechanical properties of CNT microstructures would rival those of silicon, notwithstanding the high electrical and thermal conductivity of CNTs. Microstructures made from such materials would enable major advances in microsystems technology (MEMS).

Unfortunately, although CNT forest growth is well-established, CNT forests are typically only 1-5% bulk density, and therefore the bulk properties of the CNT forests are far below those of individual CNTs. Therefore, to increase the robustness of CNT forests, post-processing methods have been developed. One approach is to densify CNT forests after growth, and there are two approaches to densification: use of direct mechanical forces by compressing, shearing, and/or rolling CNT forests during or after growth [11-14]; or use of capillary forces by submerging CNT forests in a liquid, and subsequently evaporating the liquid [15-19]. Another approach is to infiltrate CNT forests with a polymer or ceramic matrix material thus creating a nanocomposite microstructure, via a liquid-phase method such as capillary infiltration epoxy [20], or a vapour-phase approach such as CVD of a polymer or ceramic [21-23]. While the resulting composites have shown much greater mechanical properties than as-grown forests, direct infiltration of polymers is not suitable for small CNT microstructures due to forces exerted by the flow of the uncured polymer; and coating the CNTs may not be desirable because it renders the CNT surfaces inaccessible for electrical contacting or further processing.

We introduce a new process for fabricating robust CNT microstructures for use in microsystems, wherein CNT forest microstructures are densified by self-directed capillary action. In the new self-directed capillary densification process, the densifying liquid is evaporated from a reservoir and condensed onto the CNT forests, as shown in Fig. 1a. This condensation method enables the application of minute amounts of densifying liquid to the substrates, which infiltrates each structure independently and results in individual densification of each structure without the formation of capillary bridges. Within each structure, the CNTs locally aggregate by the elastocapillary mechanism [24], and coupling among the aggregating CNTs within each structure causes the entire microstructure to densify (Fig. 2a-d). As opposed to the immersion method that subjects the CNT pillars to lateral forces due to the formation of capillary bridges, the condensation method preserves much more delicate and higher aspect ratio structures (Fig. 2e-f). In addition, our new method allows use of a wide range of densifying liquids, including highly viscous liquids that result in a higher densification factor. In a separate publication, we recently showed that the self-directed densification method also facilitates design and fabrication of unique 3D CNT microstructures[25], and that densified CNT microstructures can be used as master molds for polymer replication. Herein, we develop a set of design criteria for fabrication of robust CNT micropillars by this new method, and demonstrate the mechanical and electrical properties of the micropillars.

## 2. Methods

CNT forest microstructures are grown on thermally-oxidized (100) silicon wafers, using a supported catalyst layer of 10 nm Al2O3 and 1 nm Fe deposited by e-beam evaporation [9]. The catalyst layers are patterned by a lift-off process using photoresist (SPR 220) and ultrasonic agitation in acetone. Next, CNTs are grown in a horizontal tube furnace (22 mm inner diameter, 300 mm heated length) at atmospheric pressure, with flows of 100/400/100 sccm C2H4/H2/He, at 775 °C. The growth time is typically 0.5 to 15 minutes depending on the desired forest height. Optionally, the adhesion of the CNTs to the substrate is enhanced by rapidly cooling the substrates in the growth atmosphere immediately after conclusion of the programmed growth time [13]. In this case, the furnace hood is raised with the carbon source still flowing. In some experiments, the CNTs are plasma etched prior to densification, using 10 sccm O2 at a pressure of 75 mTorr and 80 W power, for 30 seconds. This removed the tangled "crust" at the top of the forest (see Fig.6).

For the capillary densification process, the silicon substrate with CNTs is placed on an aluminum mesh (McMaster-Carr 9232T221) and inverted over a large beaker (1 L) containing a small amount of liquid (e.g., 20 mL acetone), as illustrated in Figure 3. The aluminum mesh is covered with a lab towel (e.g., a polypropylene/cellulose composite wipers, Fisher Scientific) to absorb droplets of liquid. The beaker is placed on a hotplate set to 175 °C for acetone and 300 °C for glycerol. This setup produces an upward vapor flux and maintains a lower substrate temperature that drives condensation of the liquid onto the substrate. For basic process control, we leave the substrate in place until the liquid has fully evaporated, which typically takes less than 10 minutes. The position of the sample was controlled carefully so that the samples were never "pooled" by excessive vapor condensation. This may also be addressed by creating a chamber with a cold stage that operates at sub-atmospheric pressure and controls the temperature of the sample over large areas to reliably control the condensation. The results shown in this paper are achieved using acetone and glycerol; however, 2-propanol and water have also been used with the same setup. The entire setup was placed under a snorkel vent or inside a fume hood.

In order to quantify the improvement in the density of the structures, we define a parameter called the Densification Factor (DF). The DF is defined as the ratio between the cross-sectional area of the microstructure before and after densification. Top-view SEM pictures are used to calculate the DF, by comparing the area of the catalyst to the top cross-sectional area of the structure after densification. The cross sectional area before densification is assumed to be the same as the area of the catalyst. The measurements are made using INFINITY ANALYZE software.

**3. Results and Discussion**

Figures 2a-d show typical CNT micropillars before and after densification using the self-directed capillary method. A detailed description of the process parameters and experimental setup (Fig. 3) used for these experiments can be found in the experimental section at the end of this manuscript. As shown in the close-up on the right hand side of this figure, the densification process results in a substantially higher CNT bulk density. Figures 2e and 2f compare micropillars densified using the condensation and immersion methods. As can be seen in these figures, immersion densification tends to damage closely spaced pillars, whereas the self-directed method allows the structures to densify individually.

In what follows, we present a set of guidelines for the design and fabrication of robust densified microstructures. We will address a number of critical issues for fabrication of robust CNT microstructures by capillary densification, such as the formation of internal voids and changes in the cross-sectional shape (e.g. wrinkles in Fig. 2), and the dependence of the densification factor on CNT pillar size and spacing. By variation of the CNT growth and densification process parameters, we demonstrate tuning of the densification factor, and show that densified CNT pillars made by our process have mechanical stiffness and electrical conductivity more than 1000-fold higher than as-grown CNT forests.

**3.1 Control of void formation**

Elastocapillary aggregation occurs when an ensemble of slender beams is withdrawn from a liquid, and the resultant capillary forces between the liquid and solid bring the beams closer together. Via immersion in liquids, this phenomenon has been studied for the aggregation of wet hair [24], nano- and microfibres [26, 27], polymer microstructures [28], and CNT forests [15, 29, 30]. If the capillary forces are stronger than the elastic restoring forces for the deformed beams, the beams will aggregate; and, in the case of micro- and nanostructures, surface forces can make this aggregation stable after the liquid evaporates. When the structures are distributed over a large area, e.g., a non-patterned CNT forest, the contraction of the forest during the densification together with the adhesion of the CNTs to the substrate results in the formation of randomly distributed voids. The final structure then resembles an open-cell foam. When the structures are pre-patterned into smaller areas, e.g., an array of CNT microstructures, each microstructure may have internal voids, or each structure may form a single aggregate. Although the formation of voids can be useful for certain applications such as trapping of biological cells [17], void-free structures are prerequisite for achieving microstructures with a high CNT density.

This "phase boundary" between formation of voids and formation of single aggregates depends on the dimensions of the CNT microstructures along with the diameter and density of the CNTs. We applied analytical models of elastocapillary aggregation to derive the geometrical requirements to form void-free CNT microstructures, and show that results from the self-directed capillary process achieve good agreement with this theory.

We consider a CNT forest to be an array of vertical beams that are rigidly connected to a substrate, having radius $R$ (5 nm), spacing $d$ (averaging 100 nm) and Young's modulus $E$ (1 TPa).

Following *Py et al.* [26], $L_I$, denotes the distance between the substrate and the point where two adjacent CNTs are pulled together (Fig. 1),

$$L_I = \left(\frac{9}{2\cdot(\pi-2)}\right)^{1/4} \sqrt{d \cdot \sqrt{\frac{E\cdot\pi\cdot R^3}{4\cdot\gamma}}}. \tag{1}$$

Here, $\gamma$ is the surface tension (0.025 N/m for acetone). For these model parameters $L_I$ is approximately 400 nm. If the length of the CNTs exceeds $L_I$, this paring process will continue successively to form a hierarchical structure comprising $N$ CNTs which ultimately contact at a "sticking distance" $L_S$ above the substrate. For the array of CNTs, the sticking distance is

$$L_S = L_I \cdot \left(\frac{\beta^2 \cdot (\pi-2)}{2\sqrt{\pi}3^{1/4}(2-\sqrt{2})}\right)^{1/4} \cdot N^{3/8}. \tag{2}$$

Here, $\beta$ accounts for the lattice geometry and is set equal to 0.5 (see [26]). In practical terms, an arrangement of $N$ CNTs must be longer (taller) than $L_s$ in order to form a single aggregate, without internal voids. The cluster size $N$ can be determined using the cross-sectional area of the CNT microstructure and the areal density of the CNTs which is known from the CNT diameter and the mass density of the forest [31].

Alternatively, *Zhao et al.* [15, 29, 30] derive the sticking distance as

$$L_S^3 \sim (N-1)\cdot R^2 \cdot d^2 \cdot \left(1-\sqrt{R^2/d^2}\right)^2. \tag{3}$$

Finally, *Journet et al.* reported a study of elastocapillary coalescence of non-patterned CNT forests that were wetted with a water droplet resulting in the formation of CNT "huts" with diameter $\zeta$ [29],

$$\zeta \sim \sqrt{\frac{\gamma \cdot L^4}{E\cdot R^3}}. \tag{4}$$

These analytical methods can be applied to predict the formation of voids during capillary densification. In order to validate the applicability of the elastocapillary models for design of CNT microstructures for densification, we performed a thorough dimensional study using the condensation process, using cylindrical CNT microstructures ("pillars") of varying diameter ($D$) and height ($L$). From this data, we plotted the "phase transition" between formation of a void-free densified structure, and formation of a structure with internal voids. As expected, the pillar height ($L_S$) above which micro-voids are observed increases with pillar diameter. Relatively short and wide pillars exhibit internal voids as the capillary forces cannot overcome the elastic strain necessary to bring all the CNTs together without formation of micro-voids. Relatively tall and narrow pillars form single aggregates because less elastic deformation is required to bring the CNTs together, and because the CNTs have greater length to facilitate the hierarchical elastocapillary pairing process.

As illustrated in Fig. 4a, the analytical models fit the experimental data remarkably well for CNT pillars with $D > 200$ μm. Several phenomena could explain the remaining discrepancies between our measurements and the above models. The models described above have been derived for ideally straight and uniformly spaced rods that contact only after action by capillary forces. However, as shown in Fig. 2, CNTs within a forest have significant waviness, and frequent contact. Second, in some cases the CNTs partially detach from the catalyst pattern as they are pulled together, which is not taken into account in the above models. Nevertheless, models that treat the CNTs as elastic beams are a useful guide for choosing the dimensions of CNT pillars for robust densification.

Another important consideration is the tangled "crust" layer on the top of CNT forests, which forms as the CNTs self-organize at the start of the growth process [32, 33]. Because of its tangled morphology, the layer constrains lateral movement of the CNTs during densification. Therefore, we repeated the parametric study of aggregation, with structures where the crust layer was removed by plasma etching after growth. As shown in Fig. 4B, the foam/no-foam boundary for the crust-free structures agrees with the elastocapillary models from a diameter of 150 μm onward. This improved correlation is logical since the crust layer introduces a constraint that is not taken into account in the model.

*3.2 Control of the densification factor*

After defining geometric bounds for void-free densified microstructures, we now investigate how to control the packing density of CNTs by manipulating the balance between elastic and capillary forces during the densification process. While ideally the CNTs would aggregate in perfect closely packed clusters, the tortuosity of the CNTs within the forest and the presence of the crust prevent perfect packing. Further, the diameter and initial packing density of the CNTs determines the elastic resistance to densification, and the surface tension of the liquid and the dynamics of liquid delivery may affect the strength of the capillary forces. These variables can be manipulated by the CNT pattern design, the CNT growth parameters, and the densification process parameters.

First, we study how the densification factor (DF) depends on the size and spacing of the CNT microstructures. Figure 5 shows the DF for glycerol densification as a function of the spacing for different pillar diameters and lengths, for both square and circular cross sections. Within measurement error, the DF is invariant with the spacing. Small variations could arise from the effect of pillar spacing on the amount of available area for liquid condensation which affects the infiltration dynamics, and effects of the microstructure diameter on the CNT areal density. The latter topic is currently under investigation. The DF is ≈5 for 300 μm diameter pillars which form internal voids, yet is 15-20 for 100 μm and 30 μm diameter pillars which form a single aggregate. Further, differences in the strength of CNT-substrate adhesion can affect the DF. This is due to the fact that as the pillars get densified, some nanotubes around the edge may not be long enough to get tightly bound to the central cluster, hence preventing ideal shrinkage. If these nanotubes are detached from the substrate due to capillary forces, further shrinkage may occur. The adhesion of the CNTs to the substrate is therefore another important parameter for controlling densification.

Now for a fixed pillar size and spacing, we control the densification factor by varying the structural characteristics of the CNT micropillars. First, the diameter of the CNTs is controlled by the catalyst nanoparticle size. During the catalyst annealing step, the Fe film breaks into nanoparticles which then seed the CNT growth. Thus, we found that annealing the catalyst in He without $H_2$ yields to larger particles and hence MWNTs with average diameter of ≈20 nm instead of ≈10 nm. As illustrated in Fig. 6, CNT micropillars made out of 10 nm diameter CNTs yield a higher densification than 20 nm diameter CNTs. This is because the elastic stiffness of the CNTs is proportional to $D^3$, whereas the areal density is proportional to $1/D^2$.

Fig. 6. also shows the influence of the crust layer on the densification process. Removing the crust layer by brief $O_2$ plasma etching (see experimental section) after growth enables more slip among the CNTs during the densification process, and thereby results in higher densification and more uniform radial contraction. Figure 6 shows the influence of the plasma etching step on round pillars with a 100 μm diameter using acetone. The plasma etch increases the densification factor by 2-4 fold, and results in a smoother contraction. The highest densification factor of approximately 30 is obtained with small diameter CNTs that were plasma etched. Taken

together, the influence of the pillar diameter, spacing, CNT diameter and crust layer show that the densification factor can be controlled from 4 to 30, which is a packing fraction of approximately 6% to 48% compared to an ideal packing fraction of 100% defined for hexagonally-arranged cylinders.

*3.3 Control of cross-sectional shape*

As densification occurs, the individual CNTs in the pillars come closer together which results in shrinkage in the cross sectional area of the pillar. The shrinkage is non-ideal in most cases and results in the change of the cross sectional shape. For instance, as the structures shrink, the side walls fold, resulting in wrinkled pillars as shown in Fig.2 and Fig. 6. The wrinkling of the structures could be induced by a difference in the amount of densification between the CNTs on the periphery and the core of the forest. If we assume that the CNTs on the periphery of the forest densify less (because they are more entwined or have a higher density), these sidewalls will need to wrinkle in order to adjust to the higher densification in the core of the forest. This assumption is supported by Fig. 6, which shows that after plasma treatment (which removes the top crust layer but also the outer CNT of the forest) the formation of wrinkles is greatly reduced. The latter emphasizes the advantage of the plasma treatment in order to obtain uniform microstructures by densification.

Further, we found that polygonal pillars form star-shaped cross sections after densification as illustrated in figure 7a. The number of points and arrangement of the star is dictated by the number of corners in the catalyst pattern, although as more corners are added, some points are less clearly defined. Further, as corners of the catalyst pattern get more rounded (7c), the corners in the cross section of the densified pillars also get more rounded. Removing the crust by plasma etching (7b, 7d) renders the effect of corners less pronounced as it smoothes the outline of the cross section. Finally, careful design of the catalyst pattern allows counter-acting the formation of star-shaped structures. Fig. 7e illustrates how concave structures accentuate the formation of pillars with a star-shaped cross-section, while convex catalyst patterns can cancel out this effect and result in pillars with a square cross-section after densification.

*3.4 Interactions among micropillars and the formation of hierarchical textures*

Understanding the limits to the spacing of microstructures is important for fabrication of, for example, probe needle arrays and smart surfaces. By studying the behavior of closely-spaced CNTs we determine these geometric limits and discuss the formation of hierarchical textures from pillars that contact one another during processing.

To do so, various pillar arrays are investigated, and "phases" are defined that describe the overall pillar arrangement as illustrated in figure 8. It is important to note from figure 8 that as-grown CNT pillars are not always perfectly straight. In the CNT fabrication process used in this paper, the CNTs micropillars are limited to aspect ratios of 8, above which the pillars start to bend due to growth variations. For an in-depth discussion of pre-densified arrangement of the pillars we refer to an earlier publication [33]. In what follows we will only discuss how the densification process changes the arrangement or phase of the pillars. As we will see, the straightness of the as-grown CNT pillars affects their straightness after densification, and their proximity affects how the pillars interact during densification.

We will use the same definitions of "phases" to describe the organization of the micropillars as introduced in [33]:

- Phase 1: Micropillars grow straight vertically and have tip deflections of less than its own diameter. Adjacent pillars do not touch each other.

- Phase 2 : Micropillars have tip deflections of at least one pillar diameter, yet the pillars are still self-supporting (i.e. tips of adjacent microstructures are not touching).

- Phase 3: Micropillars have tip deflections of at least one pillar diameter, and adjacent CNT pillars are touching each other.

- Phase 4: Micropillars are entwined and supporting each other in order to grow vertically, or stay vertical after densification.

- Phase 5: Micropillars fail to grow vertically or stay vertical after densification, forming a tangled arrangement of CNT bundles.

As shown in Fig. 2e-f, condensation based densification damages microstructures less than immersion based processes. However, it is obvious that for instance pillars which are touching each other after growth (phase 3), will interact with each other during the densification process. Typically, menisci are formed that couple capillary forces among these structures, or between structures and the substrate. This process aggregates pillars or pulls them down, typically resulting in phase 4 or 5 (see Fig 8). This effect is studied in more detail in figure 9 which shows the phases before and after densification of CNT pillar arrays with diameters of 5, 10, 30, 100 and 300 μm and spacings of 5, 10, 30 and 100 μm. The first column of Fig. 9 shows the phases before the densification process, while the second columns show the phases of the microstructures after glycerol densification following the densification procedure described above. This graph allows predicting which organization the micropillars will yield after densification. From top to bottom, this graph shows that the higher the aspect ratio of the pillars, the more they tend to fall over. From left to right, the graph shows the influence of densification on the organization of the pillars. Note that the closely spaced pillars retain a slightly higher phase since they support each other, which prevents them from falling down. On the one hand, this graph defines guidelines for maintaining a perfect orientation after densification. Phase 1 is for instance always retained for pillars with an aspect ratio of 1. On the other hand, some applications may require entangled micropillar architectures as shown in figure 8, phase 3-5. Figure 9 predicts for a given diameter and spacing which phase is to be expected after densification. Note that in a few cases, pillars that were phase 2, straighten out to form phase 1 structures during densification.

*3.5 Mechanical properties of densified CNT micropillars*

Due to their higher packing density, the densified CNT micropillars have a significantly greater mechanical stiffness than as-grown CNT forests. We quantified this increase by performing axial compression tests of identical arrays of cylindrical micropillars (100 μm diameter as-grown, 200 μm height), before and after densification. The pillars were plasma etched before densification to improve the uniformity (see above), and the densification factor was approximately 17. The densified pillars are shown in Fig. 10a. The structures were tested in a custom-built micro-compression testing machine using a tapered steel tip with tip diameter of 0.55 mm fixed in series to a load cell (Futek) which is rigidly mounted to a linear stage. The tip cross section is chosen to test seven cylindrical CNT microstructures in parallel. The test was performed in position control mode where compression proceeds to ≈25% strain (50 μm). The Young's modulus is calculated using the slope of the unloading curve and the measured total area of the tested structures.

Based on the load-displacement curves shown in Fig. 10b, we calculate that densification increases the Young's Modulus of the micropillars from 1.56 MPa to 2.24 GPa. The latter value is comparable to typical microfabrication polymers including SU-8 (E=2-4 GPa [34]) and PMMA (E=2-5 GPa [35]). Notably, while the reduction in cross-sectional area is 17, the modulus increases by a factor of more than 1400. The great increase in the modulus can be attributed to the change in collective loading mechanism of individual tubes within the forest which is caused by the reduction in CNT-CNT spacing [21]. Briefly, due to their waviness, the CNTs within the forests can be modeled as a number of springs in parallel, which support one another at periodic contact points. Thus individual CNTs within a forest are not baring pure compression (unlike a rod under compression), instead they are mostly subject to bending and shearing loads. The CNT-CNT spacing changes the boundary conditions on the individual sections of the springs defining new critical lengths for these spring sections (similar to springs with variable winding diameter and pitch or cantilevers with variable lengths), thus modifying their collective stiffness. Because the stiffness scales with the cube of the critical length of the individual CNT sections, densification causes a disproportionate increase in the modulus. The ability of these CNT structures to withstand large compressive strains (exceeding 25%) and high temperatures (exceeding 600 C in air [36]) makes them attractive elements for MEMS devices operating in harsh environments.

*3.6 Electrical properties of densified CNT micropillars*

In addition to being mechanically robust, the densified CNT micropillars can be electrically integrated in microsystems. We demonstrate this by growing CNT forests on patterned TiN electrodes, followed by densification and electrical characterization. The TiN electrodes were deposited by sputtering and patterned by photolithography before the catalyst patterning step. The catalyst layer consisting of 10nm $Al_2O_3$ and 1 nm of Fe was subsequently deposited by e-beam evaporation on top of the TiN electrodes through a second layer lithographic step followed by CVD growth as described in the experimental section (see figure 11a). The structures are tested by placing one probe on the TiN electrode pad while the second probe is in contact with the CNT structure from the top.

Fig. 11b compares two-point probe measurements of an individual cylindrical CNT micropillar (D = 40 μm, L = 300 μm), before and after densification. We find that the presence of the 10nm $Al_2O_3$ dielectric layer does not prevent the formation of Ohmic electric contact between the CNT structures and the bottom electrodes. The equivalent resistivity of the individual structure shown in Fig. 11a decreased from 23 Ω-cm to 9.7 mΩ-cm due to densification, which is a factor exceeding 2300:1. Although the resistivity measured in this work includes a contribution from the contact resistance between the CNTs and the probes and the series resistance of the $Al_2O_3$ layer.

As discussed for the mechanical stiffness, the enhancement in electrical properties cannot be correlated only to the reduction in cross-sectional area. The significant reduction in resistance is probably due to: (1) the decrease in contact resistance between the CNTs at the bottom of the structures and the TiN electrode pad, because the CNTs are pulled down onto the TiN electrode as they are densified, thus increasing the effective contact area; (2) improved lateral conduction between CNTs due to densification, which establishes additional flow paths around critical faults in individual CNTs; and (3) better contact with the top probe which is enabled by the higher stiffness and robustness of the densified pillars. While the as-grown pillars are very fragile and might have been poorly contacted during the measurement, we can still conclude that self-densification of CNTs capillary action is significantly advantageous for both the mechanical and electrical properties of CNT microstructures. The contact resistance between as-grown CNT forests and bottom electrodes generally prohibits their application in electrical interconnects [37, 38], needle probes [39] and other microsystems. Based on the measurements shown above, capillary self-assembly could become a key technology to overcome these challenges.

**4. Conclusions**

We demonstrated that self-directed elastocapillary densification of CNTs enables scalable fabrication of robust CNT microstructures from low-density CNT forests grown by CVD. Compared to previous methods of densification by immersion, our new approach allows densification of closely-spaced structures without undesirable aggregation. Moreover, new insights in this process have been gained by combining analytical modeling of elastocapillary densification with experimental investigation of both the CNT forest geometry and morphology. For instance, the forest aspect ratio, spacing, CNT diameter and plasma treatment all influence the densification of individual pillars as well as the interaction between closely spaced pillars. These new insights resulted in a set of design guidelines for control of the densification factor, cross-sectional shape, pillar straightness and the formation of hierarchical textures. Finally, this process is compatible with microfabrication via direct CNT growth and densification on patterned electrodes. Accordingly, we show an enhancement of three orders of magnitude in both the Young's modulus and electrical resistivity of CNT micropillars after densification. Capillary self-assembly could therefore become a key technology for the incorporation of otherwise fragile CNT forests in microdevices.

**Acknowledgements**

This research was supported by the College of Engineering and Department of Mechanical Engineering at the University of Michigan, the Belgium Fund for Scientific Research - Flanders (FWO), and the Nanomanufacturing program of the National Science Foundation (CMMI-0927634).  We thank Zhaohui Zhong for use of the probe station and Samantha Daly for use of the mechanical compression tester.

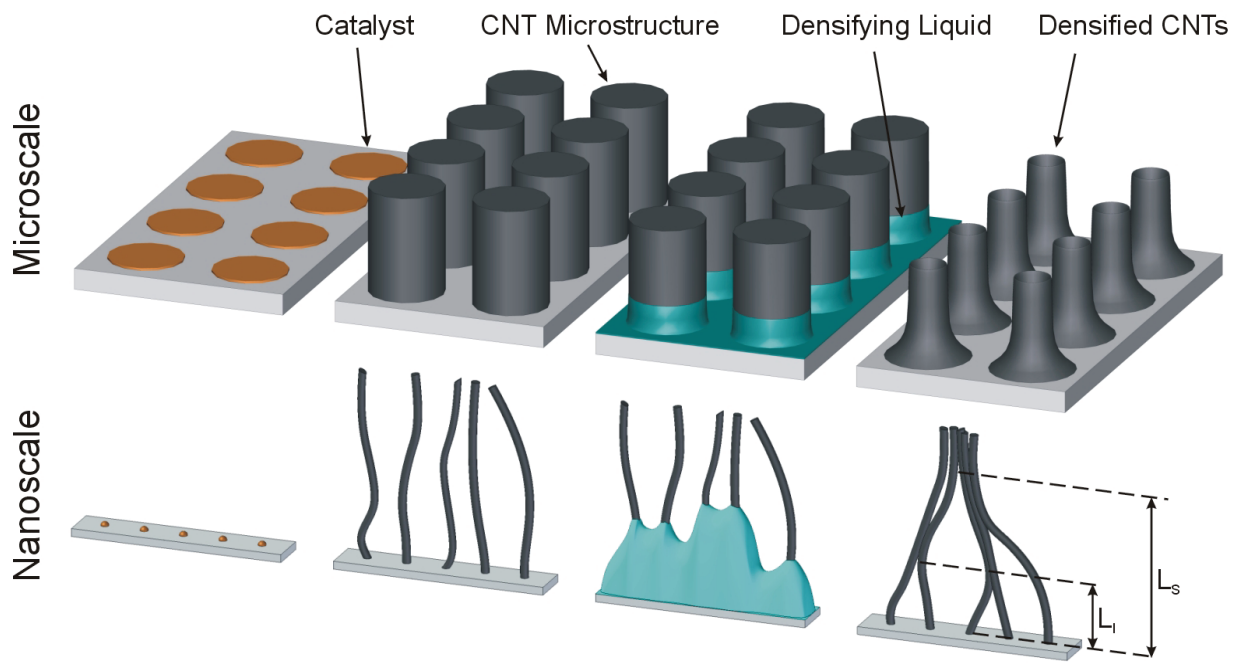

**Figure 1.** Fabrication of dense CNT micropillars by patterned growth followed by self-directed capillary densification. The densification is initiated by condensation of a solvent onto the substrate.

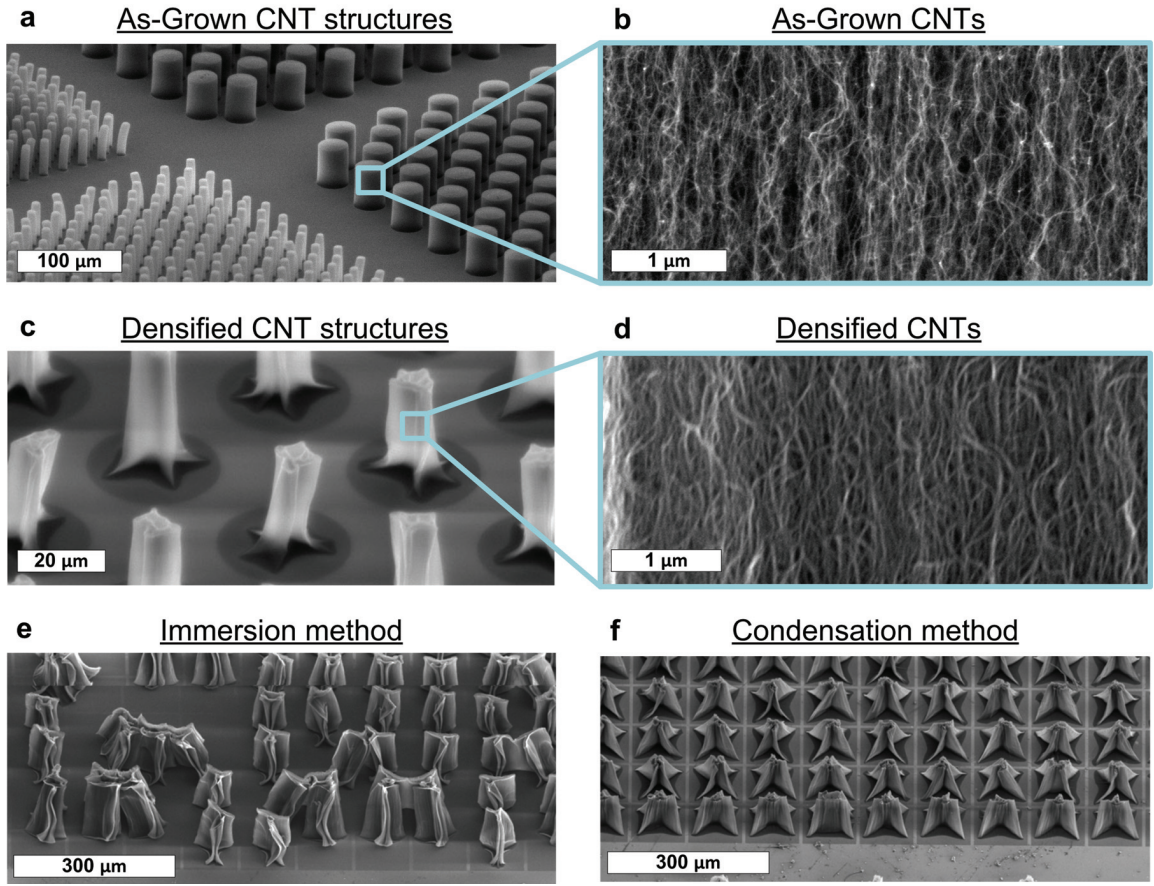

**Figure 2.** Example SEM images of cylindrical CNT microstructures before (a,b) and after densification (c,d) by the self-directed capillary method. Close-up images (b,d) show the sidewalls of the microstructures, emphasizing the significant increase in CNT density and the maintenance of the aligned topology. Comparison between vapor densification methods (e) and immersion method (f).

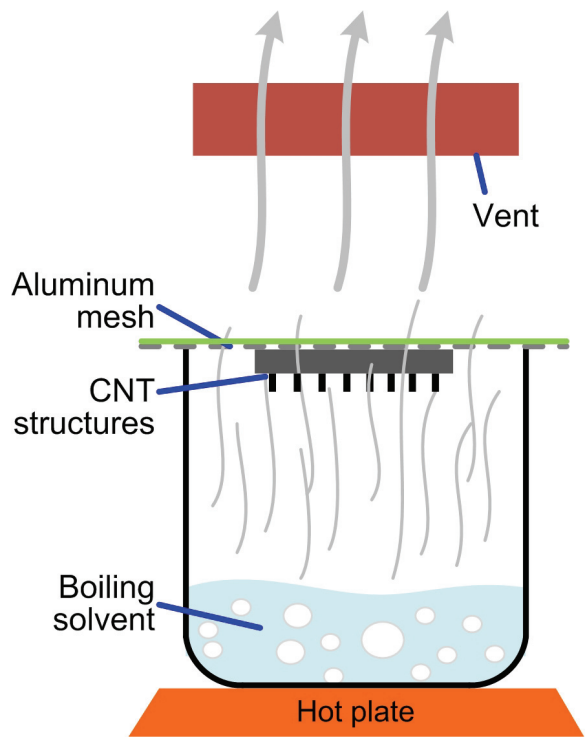

**Figure 3.** Setup for densification of CNT forests by the self-directed capillary method.

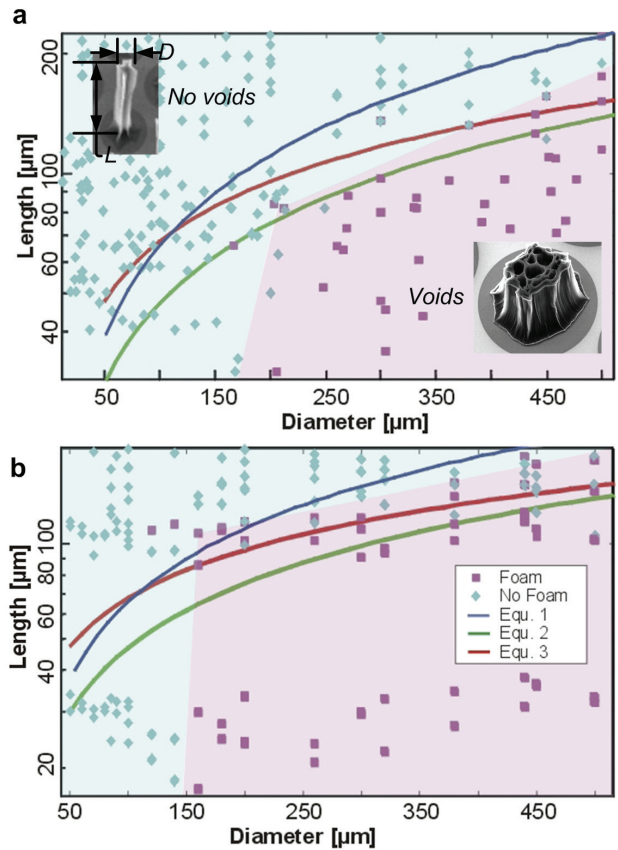

**Figure 4.** Investigation of the effect of CNT microstructure dimensions on self-directed elastocapillary densification, showing that the length (height) and diameter of cylindrical CNT forests determines whether the structures form a single aggregate, or exhibit internal voids: (a) comparison of analytical models to measurements of cylindrical microstructures; (b) comparison of analytical models to structures that were plasma etched to remove the top crust prior to densification.

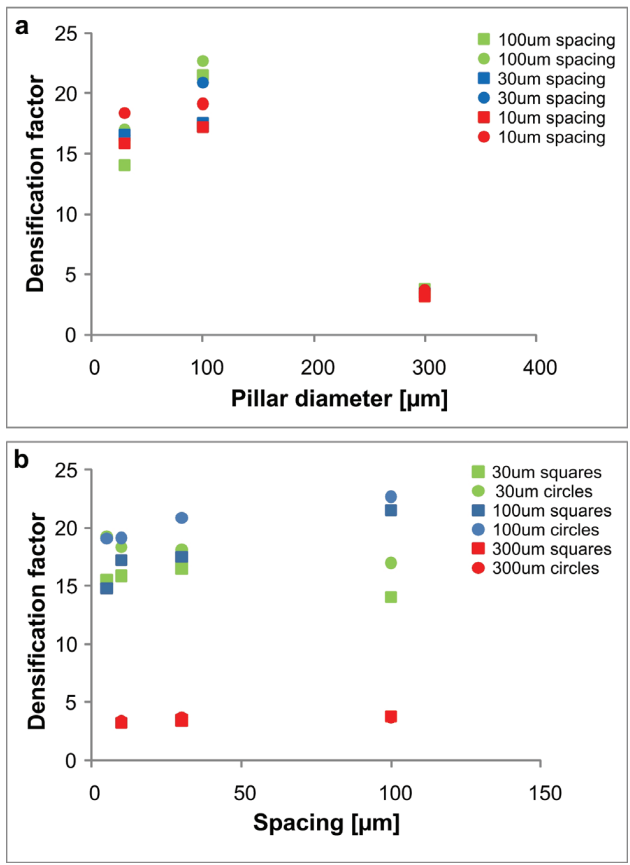

**Figure 5.** Densification factor of CNT pillars densified using glycerol as a function of (a) pillar diameter, for several spacings; and (b) pillar spacing, for several diameters. The shape of the data marker indicates the initial cross sectional shape of the CNT pillar (circle or square).

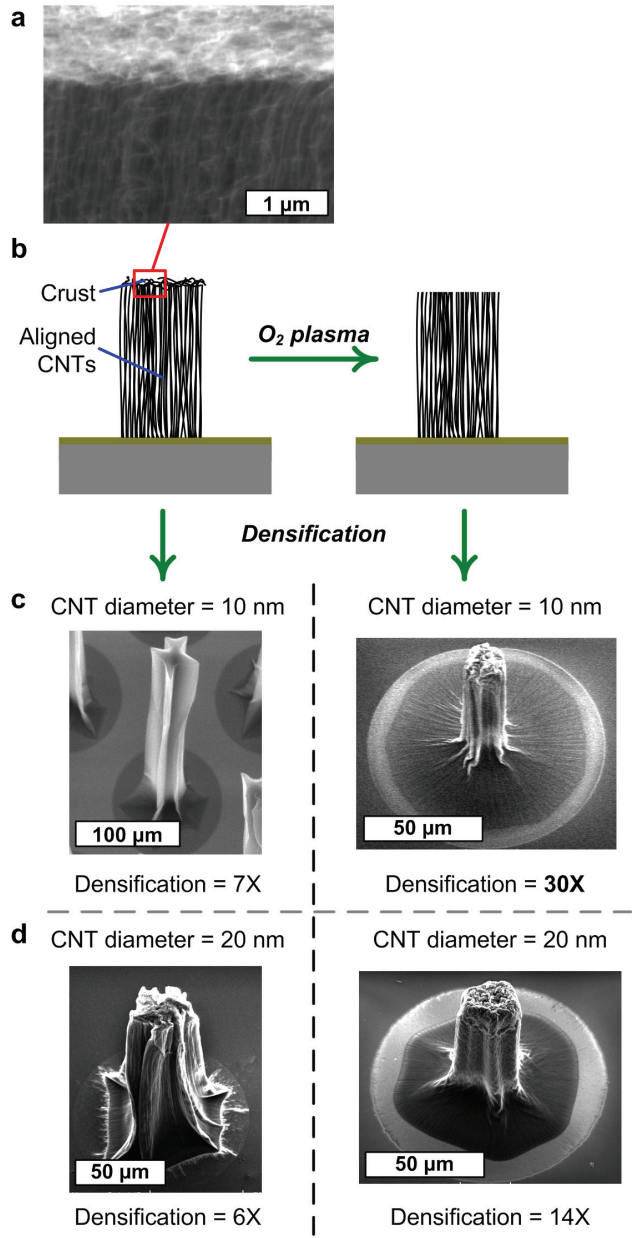

**Figure 6.** Influence of plasma etching, and the CNT diameter on the densification factor using acetone: (a) SEM image of the crust layer on top of an as-grown micropillar; (b) schematic micropillars as-grown and after plasma etching which removes the crust; (c) SEM images of micropillars after densification, with CNT diameter 10 nm; (d) SEM images of micropillars after densification, with CNT diameter 20 nm.

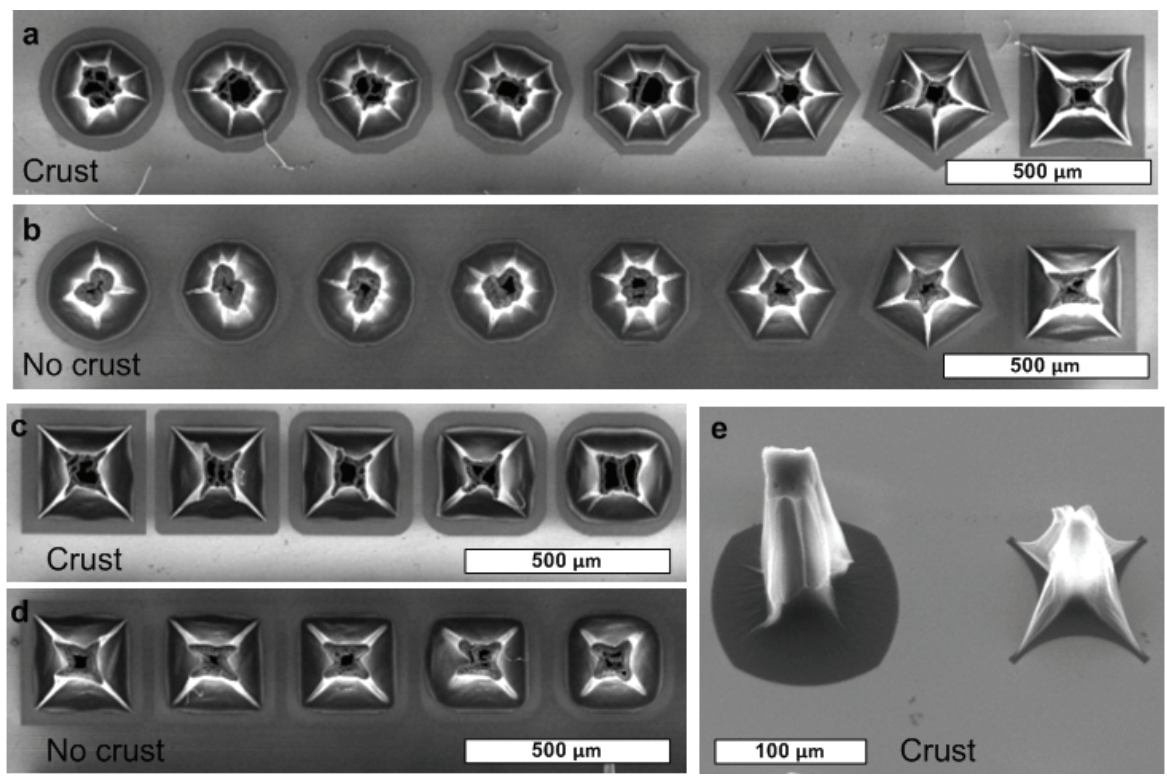

**Figure 7.** Effect of catalyst shape and crust on the resulting cross sectional shape. a) Polygonal pillars with crust b) Polygonal pillars without crust c) Square pillars with rounded corners and crust d) Square pillars with rounded corners and no crust e) Comparison of concave and convex catalyst patterns.

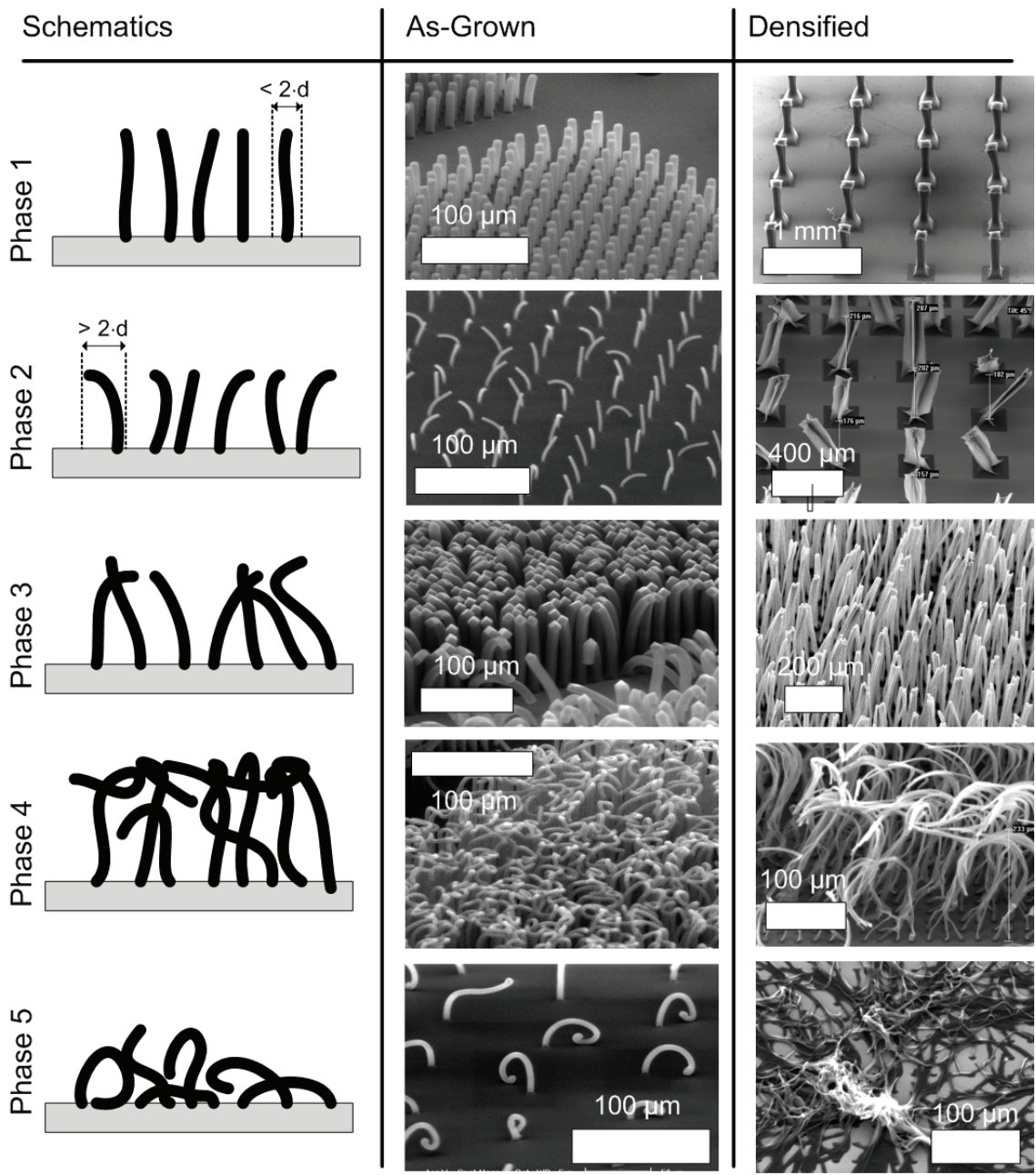

**Figure 8.** Classification of the hierarchical ordering of arrays of as-grown and densified CNT micropillars. (a) Phase 1, Straight. (b) Phase 2, Self-supporting with tip deflections of at least one-diameter. (c) Phase 3, microstructure tips are touching (d) Phase 4, microstructures grow entwined. (e) Phase 5, no vertical growth; chaos.

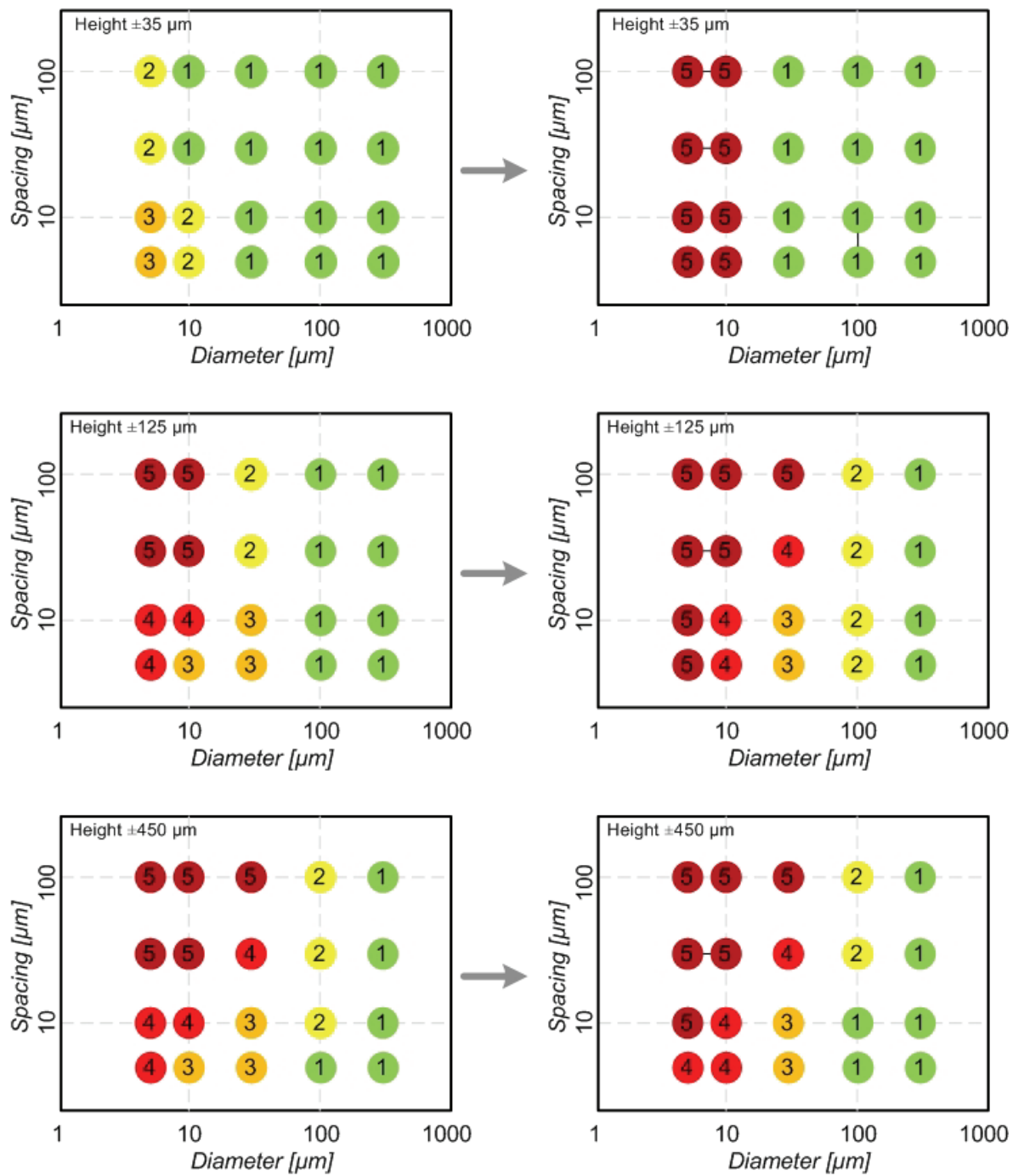

**Figure 9.** Influence of the diameter, length and spacing on the phase of round micropillar arrays before densification and after glycerol densification. No plasma treatment was applied.

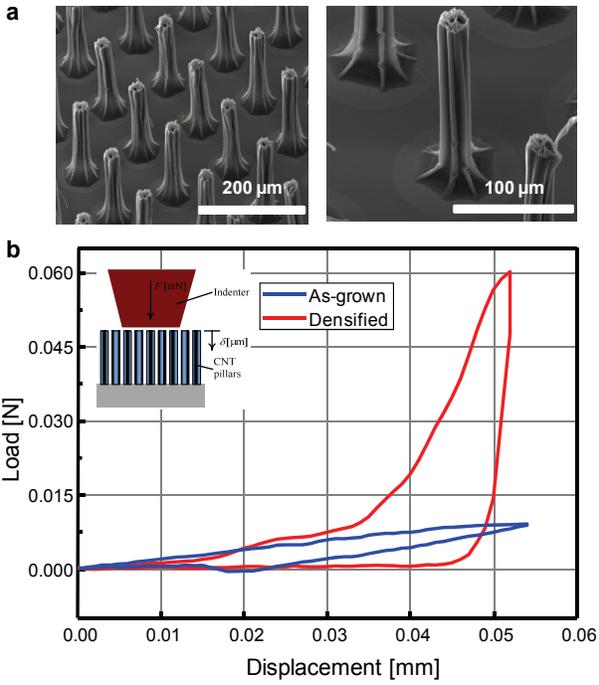

**Figure 10.** a) Densified CNT pillars with plasma treatment. B) Load-displacement curves measured on 7 micropillars in parallel.

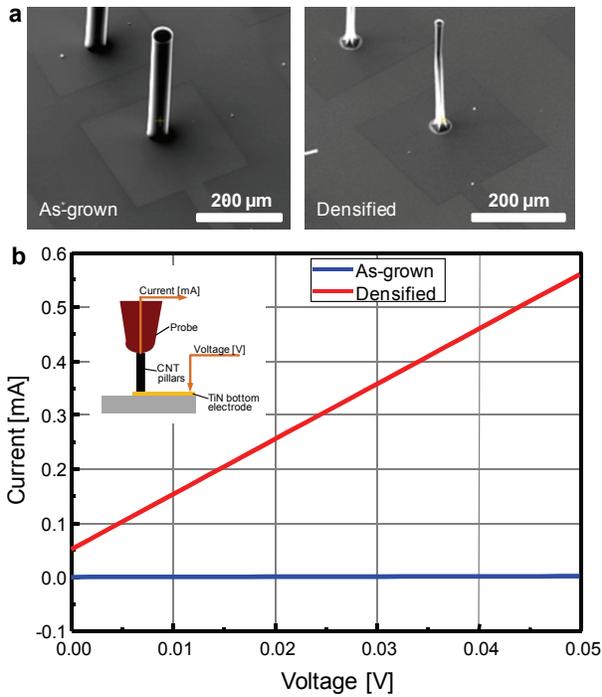

**Figure 11.** a) Undensified (left) and densified (right) CNT micropillars on TiN electrodes. b) Comparison of two-point probe measurements of the CNT micropillar resistance (D = 40 μm, L = 300 μm), before and after densification.

## References

1. Dresselhaus, M.S., G. Dresselhaus, and P. Avouris, eds. *Carbon Nanotubes: Synthesis, Structure, Properties, and Applications*. Topics in Applied Physics. 2001, Springer.
2. Harris, P.J.F., *Carbon Nanotube Science - Synthesis, Properties, and Applications*. 2009: Cambridge University Press.
3. Javey, A., et al., *Ballistic carbon nanotube field-effect transistors*. Nature, 2003. **424**(6949): p. 654-657.
4. Fischer, J.E., et al., *Magnetically aligned single wall carbon nanotube films: Preferred orientation and anisotropic transport properties*. Journal of Applied Physics, 2003. **93**(4): p. 2157-2163.
5. Nihei, M., et al., *Electrical properties of carbon nanotube bundles for future via interconnects*. Japanese Journal of Applied Physics, 2005. **44**(4A): p. 1626-1628.
6. Tong, T., et al., *Height independent compressive modulus of vertically aligned carbon nanotube arrays*. Nano Letters, 2008. **8**(2): p. 511-515.
7. Gwinn, J.P. and R.L. Webb, *Performance and testing of thermal interface materials*. Microelectronics Journal, 2003. **34**(3): p. 215-222.
8. Fan, S.S., et al., *Self-oriented regular arrays of carbon nanotubes and their field emission properties*. Science, 1999. **283**(5401): p. 512-514.
9. Hart, A.J. and A.H. Slocum, *Rapid growth and flow-mediated nucleation of millimeter-scale aligned carbon nanotube structures from a thin-film catalyst*. Journal of Physical Chemistry B, 2006. **110**(16): p. 8250-8257.
10. Madou, M.J., *Fundamentals of Microfabrication*. 2002: CRC.
11. Hart, A.J. and A.H. Slocum, *Force output, control of film structure, and microscale shape transfer by carbon nanotube growth under mechanical pressure*. Nano Letters, 2006. **6**(6): p. 1254-1260.
12. Tawfick, S., K. O'Brien, and A.J. Hart, *Flexible High-Conductivity Carbon-Nanotube Interconnects Made by Rolling and Printing*. Small, 2009. **5**(21): p. 2467-2473.
13. Pint, C.L., et al., *Formation of highly dense aligned ribbons and transparent films of single-walled carbon nanotubes directly from carpets*. Acs Nano, 2008. **2**(9): p. 1871-1878.
14. Wardle, B.L., et al., *Fabrication and characterization of ultrahigh-volume-fraction aligned carbon nanotube-polymer composites*. Advanced Materials, 2008. **20**(14): p. 2707-+.
15. Chakrapani, N., et al., *Capillarity-driven assembly of two-dimensional cellular carbon nanotube foams*. Proceedings of the National Academy of Sciences of the United States of America, 2004. **101**(12): p. 4009-4012.
16. Liu, H., et al., *Self-assembly of large-scale micropatterns on aligned carbon nanotube films*. Angewandte Chemie-International Edition, 2004. **43**(9): p. 1146-1149.
17. Correa-Duarte, M.A., et al., *Fabrication and biocompatibility of carbon nanotube-based 3D networks as scaffolds for cell seeding and growth*. Nano Letters, 2004. **4**(11): p. 2233-2236.
18. Futaba, D.N., et al., *Shape-engineerable and highly densely packed single-walled carbon nanotubes and their application as super-capacitor electrodes*. Nature Materials, 2006. **5**(12): p. 987-994.